# Memory Management Strategies for an Internet of Things System

Ana-Maria COMEAGĂ, Iuliana MARIN, *Member, IEEE*

*Abstract*— The rise of the Internet has brought about significant changes in our lives, and the rapid expansion of the Internet of Things (IoT) is poised to have an even more substantial impact by connecting a wide range of devices across various application domains. IoT devices, especially low-end ones, are constrained by limited memory and processing capabilities, necessitating efficient memory management within IoT operating systems. This paper delves into the importance of memory management in IoT systems, with a primary focus on the design and configuration of such systems, as well as the scalability and performance of scene management. Effective memory management is critical for optimizing resource usage, responsiveness, and adaptability as the IoT ecosystem continues to grow. The study offers insights into memory allocation, scene execution, memory reduction, and system scalability within the context of an IoT system, ultimately highlighting the vital role that memory management plays in facilitating a seamless and efficient IoT experience.

## I. INTRODUCTION

THE main aspects of our lives like thinking, education, working or entertaining, were totally changed by the existence of the Internet, and now a new concept is growing rapidly, because of the diverse application domains and the progress in wireless communication technologies, namely the Internet of Things (IoT). This will have a greater impact on our lives and enable the connection of more devices than any other aspect of the digital age has accomplished thus far. In the context of the Internet of Things, "things" refer to equipment with sensing, actuating, storage, or processing capabilities [1]. These devices exhibit distinctive traits, such as limited memory, reduced battery capacity, and constrained processing power.

The focus on designing templates for creating customized monitoring and assistance scenarios reflects a broader shift towards user centric IoT systems to create personalized solutions that align with their needs [2].

A.-M. Comeagă is with the Department of Engineering in Foreign Languages, National University of Science and Technology POLITEHNICA Bucharest, Bucharest, RO 060042 (e-mail: ana_maria.comeaga@stud.fils.upb.ro).

I. Marin is with the Department of Engineering in Foreign Languages, National University of Science and Technology POLITEHNICA Bucharest, Bucharest, RO 060042 (e-mail: marin.iulliana25@gmail.com).

Our approach is in line with this emphasis on user customization and responsiveness. Moreover, IoT has a great potential to improve our society and create a smart world for people by facilitating intelligent communication among objects and devices in an economical way. Systems are designed for the management of various devices, ranging from advanced Medical and Defense Systems to household appliances, such as stoves, ovens, TVs, washing machines, and more. This places significant emphasis on the importance of IoT in enhancing Embedded Systems, as they are at the core of implementing these functions [3].

IoT devices are characterized by constrained memory and power resources, needing real-time capabilities in specific situations. Furthermore, they must be compatible with diverse hardware and incorporate efficient connectivity and security measures. The paramount challenge facing the research community is establishing efficient methods for connecting and managing this extensive array of devices. Within an IoT system, the primary research concern revolves around the structured and controlled allocation of available resources [4]. The goal of an IoT resource management mechanism is to proficiently meet the requirements of IoT devices.

There are two general categories of IoT devices: low-end IoT devices and high-end IoT devices [5]. Low-end devices are severely limited in terms of available resources. These small resource-constrained devices are not supported by all operating systems (OSs), especially the traditional ones, like Linux, so IoT cannot realize its complete potential until a universally recognized operating system is in place to facilitate the operation of these resource-constrained devices across a diverse network. Low-end devices have minimal Random Access Memory (RAM) and limited processing capabilities. These devices also necessitate real-time capabilities in situations, such as vehicular communications, healthcare systems, factory automation, and surveillance applications. Furthermore, high-end devices, such as smartphones and Raspberry Pi, boast greater processing power and energy capacity [6]. The primary goal of IoT is to establish communication with energy efficiency and reliability. Low-end IoT devices are equipped with limited memory and processing power. Therefore, selecting an appropriate lightweight operating system is crucial to meet the resource constraints of these low-end devices.

It is essential for an IoT operating system to exhibit a high degree of concurrency to facilitate the sensing operations of low-end devices [7]. Thus, the main motivation for this paper writing is efficient memory allocation in the operating systems for low-end devices. Various memory management strategies, including paging, segmentation, and virtual memory, with their advantages, disadvantages, and performance.

In the current paper is presented an IoT system with features that manage the needs of modern home assistance and monitoring. This IoT system is not just a collection of interconnected devices, it is a dynamic ecosystem that streamlines the discovery, installation, renewal, and removal of monitoring devices, to enhance home assistance. In the next section is presented the literature review regarding resources management for devices, followed by the proposed IoT system. Section 3 outlines the results after testing the system. The last section outlines the conclusions and future work.

## II. Methodology

The forthcoming IoT landscape will be capable of independently managing and executing tasks. The adoption of a component-based architecture, with a clear separation between components interacting with the IoT domain and those responding to events, is in harmony with the contemporary trend of modularity and flexibility in IoT system design. Modular IoT architectures have been explored in research, highlighting the advantages of scalability, robustness, and ease of maintenance [8]. Our choice to implement this architecture acknowledges its growing relevance in modern IoT systems.

Treating each device as an entity, configured using YAML files, echoes the industry's move towards standardized and human-readable configuration methods. Research in IoT device management and interoperability underscores the importance of defining device representations and configurations in a structured and easily understandable manner [9]. Our approach aligns with this trend and prioritizes simplicity and accessibility in device management.

The emphasis on real-time monitoring and control of various device aspects mirrors the expectations of contemporary IoT users. Recent research in IoT user interfaces highlights the importance of responsive and interactive systems that provide real-time feedback and control options [10]. Our system's capacity for real-time interaction aligns with these user-centric principles. The use of event triggers to simulate real-world scenarios is consistent with best practices in IoT system testing and validation. Research in IoT testing emphasizes the significance of scenario-based testing to ensure the reliability and robustness of IoT applications [11]. Our approach aligns with these recommendations, facilitating comprehensive testing and validation.

An ultra-dense network introduces significant computational complexity [12]. To address challenges posed by IoT low-end devices, which include resource limitations and distributed, densely populated environments, there is a substantial demand for an effective resource management mechanism within an IoT operating system [13]. The efficiency of device's resources management is firstly given by the IoT operating system. A lot of different operating systems tried to find proper solutions to satisfy the main needs of low-end devices [14]. To make this goal possible, operating systems implemented various mechanisms to ensure the appropriate operation of sensory nodes. In most cases, the IoT low-end devices work with limited battery power [15]. As a result, it is of paramount importance to offer an energy-efficient operating system. These resource-constrained devices transmit the gathered data through a communication protocol. To achieve energy efficiency, these communication protocols aim to conserve the maximum amount of energy. This energy-efficient approach must extend to protocols operating at the transport layer, MAC layer, and network layer [16].

IoT devices rely on computational capabilities to perform their sensing tasks. These limited sensing motes typically come with restricted memory and processing capacities, usually consisting of 100 kB of flash memory and 10 kB of RAM [17]. Because of this limitation, the devices must manage their memory efficiently. Furthermore, the compounding factors of densification, randomness, and uncertainty pose considerable challenges in managing resources for IoT devices. An operating system serves as the resource manager within an intricate IoT system. To address the constraints imposed by limited processing power and memory, an OS must employ a robust process and memory management mechanism. The primary goal of an IoT system revolves around conducting sensing operations and transmitting the collected data to the base station for subsequent processing. The design of communications, signal processing, data reception, data transmission, and the radio's sleep/wake mechanism must all prioritize efficiency in terms of energy consumption and communication effectiveness [18]. In IoT operating systems, data is stored, organized, and retrieved through a file system. Consequently, the incorporation of an efficient, resilient, and suitable file system is of great significance in IoT operating systems.

### A. Importance of Memory Management

Memory management within an operating system constitutes a methodology for the control and governance of RAM, which serves as primary memory [19]. Its purpose is to enhance concurrent processing, system efficiency, and memory utilization. This memory management mechanism involves the relocation of processes between primary memory and secondary memory, along with simultaneously monitoring the status of available, allocated, and unallocated memory resources.

Memory management facilitates the execution of computer programs that demand a greater amount of primary memory than what is readily accessible within the system's available free memory. This is accomplished by transferring data between primary and secondary memory. It deals with the primary memory of the system by creating abstractions that make programs running on the system perceive that they have been allocated a substantial amount of memory. The responsibility of memory management is to safeguard the memory assigned to each process from potential corruption caused by other processes [20].

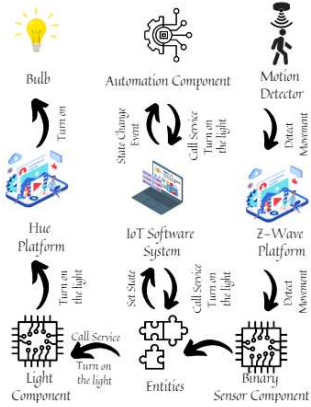

Fig. 1. IoT system components

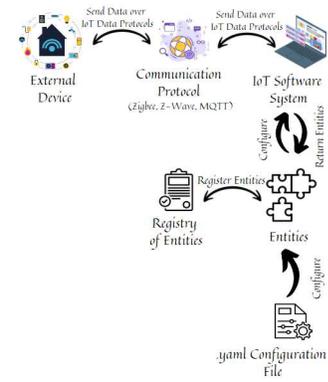

Fig. 2. Entity management within the IoT system

Memory management facilitates the sharing of memory spaces among processes, allowing multiple programs to occupy the same memory location, albeit one at a time. Moreover, it encompasses primary and secondary memory control. The memory management module is responsible for the allocation and deallocation of memory for programs. Within an operating system, various storage management functions are performed, including monitoring the utilization status of storage media to determine which memory segments are in use and which are not. When a process requests memory, an operating system assists the allocation process, and if a process no longer requires memory, the memory is released. In a multi-programming environment, memory allocation tasks are handled with the assistance of the operating system.

*B. Methods to Allocate Memory*

Memory management provides various processes and threading options for memory allocation and deallocation strategies [21]. Operating systems offer two prevalent memory allocation approaches: static and dynamic. Static memory allocation occurs during compile time, necessitating the allocation of all the memory required by a program, applications, or variables at the outset of execution. Even if a significant portion of the allocated memory remains unused during a specific program instance or for variables, it cannot be repurposed for other tasks [22]. In Dynamic Memory Allocation (DMA), memory allocation occurs during runtime, and it is assigned whenever a program, application, data, or variable requests a specific number of bytes as needed. Once memory is assigned and is no longer needed by a program, application, it can be made available for other purposes.

III. METHODOLOGY – DESIGN AND CONFIGURATION OF AN IOT SYSTEM

In this chapter, we will delve into the methodology employed for designing and configuring an IoT system. The IoT software system is implemented using Python as the backend and HACS (Home Assistant Community Store) for the graphical user interface (GUI). This system facilitates the discovery, installation, renewal, and removal of monitoring devices, specifically designed for home assistance. It serves as a user interface and plays a pivotal role in simplifying the design of tailored monitoring and assistance solutions.

The software solution provides design templates for creating customized monitoring and assistance scenarios. The created IoT system comprises two types of components: those interacting with the IoT domain and those responding to events. One such example is illustrated in Fig. 1, where smart bulbs from Philips are integrated using the Hue platform, as well as a motion detector that uses the Z-Wave protocol to communicate with the system.

Components that interact with an Internet of Things domain track the devices in the recipient's home and consist of a central part and module-specific logic. These components make their information available through the State machine and the Event bus event infrastructure. This event bus implements the publish-subscribe model. The event bus is used to decouple components so that one component can react to events fired from another component without them having direct dependencies between them. They only need to know the event bus. Each user can subscribe to a specific event. A user will be notified when the event they subscribe to is published on the event bus. A publisher can publish events on the event bus when something happens.

Components also register services in the Service Registry to expose device control. Components that respond to happening events provide home automation logic and involve services that perform common tasks in the user's home. Each device is represented as an entity. An entity abstracts the inner workings of the IoT system. The integrator administrator or designer extends an entity class and implements the necessary properties and methods for the type of device they are integrating as shown in Fig. 2. Configuring a new device to do using a human-readable yaml file. In addition, it can be used with all programming languages. The entity is responsible for distributing configuration to the IoT platform, redirecting configuration inputs and discoveries to devices, and collecting entities for service calls. All entities for the platform are managed and updated when necessary. When adding entities, the entity registry will be queried to ensure that entities to be added have the correct IDs. The entity registry will track entities and allow users to store additional settings for an entity. In the proposed IoT system, each device is represented as an entity. An integrator or designer extends an entity class and implements the properties and methods required for the type of device being integrated.

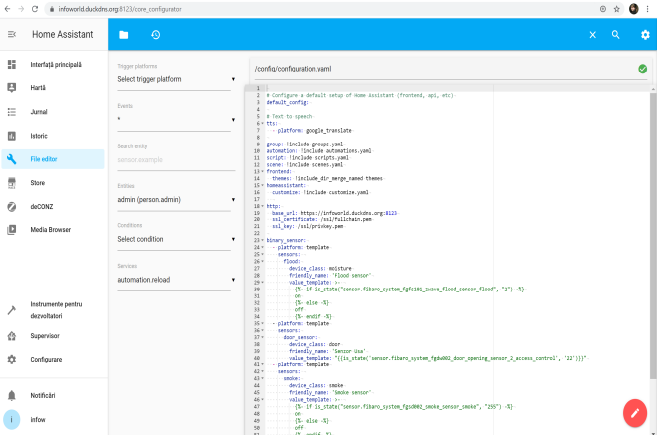

Fig. 3. YAML configuration file

Entities are configured using YAML files, which can be human-readable and language-agnostic. An entity is responsible for distributing the configuration on the IoT platform, redirecting configuration inputs and discoveries for devices, and collecting entities for service calls. Entities, which represent various aspects of devices, can be monitored, and controlled through the IoT system interface. The IoT system displays the current state of these entities. The state can be binary or numeric, depending on the entity's type. The configuration .yaml file from Fig. 3 is utilized to register new devices that will be monitored, as illustrated in Fig. 4. The managed devices arranged in rows consist of two light bulbs, a spotlight, a motion sensor (positioned under the red-colored spotlight), three switches for manual control of the light bulbs and the spotlight, three circular-shaped sensors for detecting smoke, carbon monoxide, and flooding, a panic button (located between the switch and the socket), a sensor for intrusion detection (the vertical one), and an electrical outlet. The IoT system allows users to call services for platform control, like refreshing a web page or physically turning on a light. Users can check the functionality of these services, such as in the case of light intensity or color. Custom panels in the user interface were created using the Jinja2 web template engine for Python. Events were automatically triggered to test their behavior, simulating real-world scenarios.

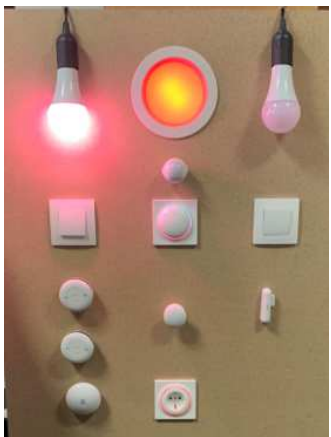

Fig. 4. Managed IoT devices in the proposed system

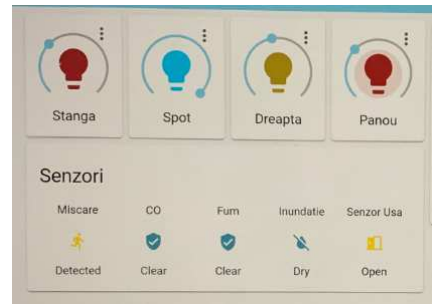

Fig. 5. Graphical interface when events are detected

The presence and opening of the intrusion sensor (door sensor) are detected by a change in the logo's color to yellow (as shown in Fig. 5).

## IV. RESULTS

The current chapter explores the automation process within the proposed IoT system. Automations are triggered by events, checked against predefined conditions, and execute services based on those conditions. Automation scenarios, such as lighting changes due to increased room temperature, were tested, as well as the ability to manually trigger automations for testing purposes. Scene management is a critical feature in the IoT system IoT system, providing users with the ability to define, customize, and control scenarios within the IoT project. These scenes encapsulate business rules, configurations, and automation instructions related to the operation of IoT devices and services in a user's home or care environment. Scene management empowers users to orchestrate a wide range of actions and device interactions, enhancing the overall functionality and convenience of the system. There are some key aspects of scene management in the IoT system. Firstly, customization, when scenes allow users to tailor the system's response to specific situations and preferences. For example, users can create scenes for different times of the day, such as morning, evening, or night, each with its own set of actions and device control instructions. This customization enables a more personalized and adaptable smart home or care environment.

Another key aspect of scene management is represented by configuration when users can configure scenes by specifying which devices or entities should be involved in the scene and what actions those devices should perform. These actions can include turning lights on or off, adjusting thermostat settings, locking doors, and more. The flexibility of scene configuration enables users to define a wide range of automation scenarios.

Scenes are triggered by events, conditions, or user-initiated actions. The tested events include sensor inputs (e.g., motion detected, door sensor, as in Fig. 5). Versatility ensures that scenes respond appropriately to the context. Scenes often involve automation, where predefined actions are executed automatically when triggered. For instance, a morning scene turns on lights when the user wakes up, enhancing comfort and convenience. Scene management in the IoT system is user-friendly, allowing users to create, modify, and activate scenes through an intuitive interface.

TABLE I
TESTED AUTOMATION SCENARIOS

| Scenario Description | Automation Type |
|---|---|
| Lighting changes due to increased room temperature | Sensor-triggered |
| Manual triggering for testing purposes | User-initiated |
| Customized morning scene | Time-triggered |
| Evening scene with device control | Condition-based |
| Complex room automation | Multiple condition-based |

Regarding integration, scenes incorporate multiple IoT devices and services, enhancing the quality of life for individuals requiring care. Scene management is scalable, adapting to the growth of a user's home environment. Both developers and users can create intricate scenes involving multiple conditions and triggers, effectively balancing energy efficiency and user comfort. Scene management is pivotal in the present IoT project's mission to deliver effective, user-centric home assistance, contributing to a safer, more convenient, and efficient living environment for IoT assistance beneficiaries. In the context of the proposed IoT system, efficient memory management plays a crucial role in ensuring the smooth operation of the IoT system module and the entire ecosystem. Memory management is not only essential for handling the software components but also for accommodating the large volume of data and configurations associated with scene management and IoT device integration.

Five scenarios have been studied in the current paper, as in Table 1. In the first scenario, lighting responds to room temperature changes as the automation system adapts to temperature fluctuations. When the temperature exceeds a predefined threshold, the system automatically adjusts the lighting. For example, if the room becomes too warm, the lights may dim to save energy. Manual testing in this scenario involves evaluating the system's response to user-initiated triggers, allowing users to activate automation processes for assessment and fine-tuning. The customized morning scene empowers users to create personalized scenes, such as "Good Morning," scheduled to trigger at specific times, like turning on the lights when the user wakes up. Evening scenes include actions like adjusting thermostat settings, locking doors, and other device controls, all aimed at enhancing user comfort.

Complex room automation scenarios involve multiple conditions and triggers, such as detecting user presence, considering the time of day, and adjusting lighting and climate control to optimize energy efficiency and user comfort.

Efficient memory management strategies tailored to the specific requirements of the IoT scenarios have been employed to handle different measurement intervals (15 seconds, 30 seconds, one minute, two minutes, and five minutes), as detailed in Table 2. In response to room temperature changes, memory management strategies adapt to different measurement intervals. For 15-second tracking, a lightweight circular buffer is utilized, continuously overwriting older data to save memory space. At 30-second intervals for more comprehensive analysis, the buffer size is extended. When measurements are taken every minute, the system stores average values, discarding older data. For two-minute intervals, temperature trends are calculated, effectively reducing the number of data points. In the case of five-minute measurements, changes are efficiently summarized.

A 24-hour study focused on an elderly individual living alone, the primary concern is their comfort, health, and safety. Temperature tracking occurs at 15-second intervals, employing memory-efficient strategies. Transitions between morning and evening and comprehensive daily analysis are implemented to ensure the resident's well-being while optimizing memory usage.

In manual testing, effective memory management is crucial. At 15-second intervals, a concise log with timestamps focuses on recent events for log manageability. Expanding the log at 30-second intervals captures more details, with pruning of older entries to prevent log inflation. Measurements every minute involve data aggregation to reduce log entries, while maintaining brevity at two-minute intervals while capturing key data. Measurements every five minutes provide a summarized overview, resulting in a memory-efficient and informative testing record.

In a 24-hour scenario monitoring an elderly resident living alone, the primary focus is their safety and well-being. During the night, the smart home system tracks room conditions, including lighting and motion sensors, with event logging every 15 seconds and pruning of older entries for efficiency. As morning approaches, temperature and lighting are monitored every 30 seconds with more detailed logs. When the resident is active, manual control devices are observed every minute for safety, reducing log entries for efficiency. In the evening, intrusion detection and temperature trends are monitored at two-minute intervals with reduced data points. Throughout the day, every five minutes, comprehensive analysis summarizes events, offering a memory-efficient and informative approach for the resident's review.

TABLE II
MEMORY MANAGEMENT STRATEGIES BASED ON TIME INTERVAL

| Scenario Description | Every 15 seconds | Every 30 seconds | Every one minute | Every two minutes | Every five minutes |
|---|---|---|---|---|---|
| Lighting changes due to increased room temperature | Circular buffer | Extend buffer | Rolling avg. | Trend analysis | Summarized data |
| Manual triggering for testing purposes | Basic log | Extend log | Aggregate data | Aggregate data | Summarized data |
| Customized morning scene | Rolling buffer | Extend buffer | Rolling avg. | Rolling avg. | Summarized data |
| Evening scene with device control | Log recent events | Extend log | Aggregate data | Aggregate data | Summarized data |
| Complex room automation | Circular buffer | Extend buffer | Rolling avg. | Trend analysis | Summarized data |

In a scenario dedicated to monitoring a customized morning scene, the objective is to ensure the resident's comfort and convenience. Early morning monitoring includes user activity and lighting status, tracked every 15 seconds using a rolling buffer for efficient parameter changes. As the morning progresses, the need for more precise monitoring arises. The system transitions to 30-second intervals with an extended buffer for detailed tracking. Daytime maintains optimal morning conditions, with one-minute intervals storing average user activity and lighting. Shifting to an evening scene with different requirements, user activity and lighting are monitored every two minutes, calculating averages for fewer data points and optimized memory usage. Throughout the day, every five minutes, comprehensive analysis summarizes user activity and lighting, providing a broader view. This ensures memory efficiency and understanding of the resident's preferences and patterns.

For evening scenes with device control, 15-second intervals capture user-initiated actions and device status changes while prioritizing recent events. At 30-second intervals, we expand the log to include detailed information but maintain pruning to manage the log's size. When measurements occur every one-minute, we aggregate data for fewer log entries. For two-minute intervals, we summarize data to optimize memory usage, and for measurements every five minutes, we provide an informative summary of the evening scenes.

In complex room automation, when measurements occur every 15 seconds, recent sensor data and automation triggers are stored in a dynamic circular buffer to capture real-time changes effectively. Extending the buffer size for 30-second measurements accommodates more intricate scenarios, suitable for monitoring rapid changes. Transitioning to one-minute intervals, the focus shifts to computing averages and trends over one-minute periods, reducing stored data points while retaining valuable insights. For measurements every two minutes, advanced data reduction techniques come into play, allowing for trend calculation and summarization over two-minute intervals, optimizing memory usage. At a five-minute interval, complex automation scenarios are summarized to highlight high-level trends and actions, instead of storing individual data points, as this provides an overview without excessive data storage.

## V. Conclusions

This study focuses on the IoT system module and its memory management role, acting as a user-friendly interface for managing IoT scenarios, with a focus on potential future directions. The proposed IoT system simplifies device integration and management, reducing overhead and memory consumption. Future work will extend to improving energy efficiency, strengthening security features, and enhancing the user experience.

As new IoT devices and technologies continue to emerge, the IoT system should remain at the forefront of integration, with future endeavors focusing on establishing partnerships and standards for seamless integration. In conclusion, the proposed IoT system, with the IoT system module as its core component, holds immense potential for revolutionizing home assistance. Continuous development and improvement, with a significant focus on memory management, will be instrumental in maintaining its effectiveness and relevance in the dynamically evolving IoT landscape.